\documentstyle[graphicx,url]{mn2e}
\footnotesize
\newdimen\minuswidth    %define @ width of minus sign for tables
\setbox0=\hbox{$-$}
\minuswidth=\wd0
\catcode`@=\active
\def@{\kern\minuswidth}
\newdimen\digitwidth    %define ! a one digit width for tables
\setbox0=\hbox{\rm0}
\digitwidth=\wd0
\catcode`!=\active
\def!{\kern\digitwidth}
\normalsize
\title[TEMPO2, a new pulsar timing package. III: Gravitational wave simulation]
{TEMPO2, a new pulsar timing package. III: Gravitational wave simulation}
\author[G. Hobbs et al.]
{G. Hobbs,$^1$
F. Jenet,$^2$
K. J. Lee,$^3$
J. P. W. Verbiest,$^{1,4}$ 
D. Yardley,$^{1,5}$
\newauthor
R. Manchester,$^1$
A. Lommen,$^6$
W. Coles,$^7$
R. Edwards,$^8$ 
C. Shettigara$^9$
\\
$^1$ Australia Telescope National Facility, CSIRO, PO~Box~76, Epping
NSW~1710, Australia \\
$^2$ Center for Gravitational Wave Astronomy, University of Texas at
Brownsville, 80 Fort Brown, Brownsville, TX 78520, U.S.A.\\
$^3$ Department of Astronomy, Peking University, 5 Haidan Lu, Beijing
100871, China \\ 
$^4$ Centre for Astrophysics and Supercomputing, Swinburne University of Technology, P.O. Box 218, Hawthorn VIC 3122, Australia\\
$^5$ School of Physics, University of Sydney, NSW, Australia\\
$^6$ Franklin and Marshall College, 415 Harrisburg Pike, Lancaster, PA 17604, U.S.A.\\
$^7$ Electrical and Computer Engineering, University of California at San Diego, La Jolla, California, U.S.A \\
$^8$ 10 James Street, Whittlesea, Vic. 3757, Australia\\
$^9$ School of Chemistry and Physics, University of Adelaide, South Australia, 5005, Australia 
}

\let\leqslant=\leq %Authors with AMS fonts and mssymb.tex can comment
\date{}
\def\sp{\mbox{\space}}

\begin{document}
\maketitle
\newcommand{\setthebls}{
%                 de-comment this line for double spacing:
%\baselineskip=20pt
}
\setthebls
\begin{abstract}

Analysis of pulsar timing data-sets may provide the first direct detection of gravitational waves.    This paper, the third in a series describing the mathematical framework implemented into the \textsc{tempo2} pulsar timing package, reports on using \textsc{tempo2} to simulate the timing residuals induced by gravitational waves.  The \textsc{tempo2} simulations can be used to provide upper bounds on the amplitude of an isotropic, stochastic, gravitational wave background in our Galaxy and to determine the sensitivity of a given pulsar timing experiment to individual, supermassive, binary black hole  systems. 
\end{abstract}


\section{Introduction}

Sazhin (1978)\nocite{saz78} and Detweiler (1979)\nocite{det79} were the first to realise that pulsar timing observations provide a powerful tool for detecting ultra-low frequency  ($f_g \sim 10^{-9}$\,Hz) gravitational waves (GWs).  The precision with which millisecond pulsars are now being timed makes it possible that pulsar timing experiments could provide the first direct detection of a GW signal\footnote{Observations of the first binary pulsar, B1913+16 (Hulse \& Taylor 1974)\nocite{ht74}, provided the first evidence for the existence of GW emission. The pulsar timing experiments described in this paper are designed to make a direct detection of GWs.}.  The Parkes Pulsar Timing Array (PPTA) project (e.g. Hobbs 2008, Manchester 2008 and references therein)\nocite{hob08}\nocite{man08} is an attempt to achieve this ambitious goal by making regular observations of 20 bright millisecond pulsars. 

Recent theoretical work (e.g. Jaffe \& Backer 2003, Wyithe \& Loeb 2003\nocite{jb03}\nocite{wl03a}) suggests that the strongest signal potentially detectable by such experiments would be an isotropic stochastic GW background caused by coalescing supermassive black holes in the centres of merging galaxies. Jenet et al. (2005)\nocite{jhlm05} showed that in order to detect this signal, the 20 PPTA pulsars will need to be timed to a precision of $\sim$0.1\,$\mu$s over a timespan of $\sim$5\,yr.   To date, the PPTA project has data spanning $\sim$3\,yr with root-mean-square (rms) residuals of typically 0.1$-$3$\mu$s, but it is expected that these residuals will significantly improve over the next few years with new observing systems and enhanced signal processing procedures.  Therefore, it is now appropriate to determine how these existing data-sets can be used to limit the amplitude of GW signals and how future data-sets will be analysed in order to detect a GW signal and determine its properties. 

Pulsar observations lead to measurements of pulse times-of-arrival (TOAs; $t_a^{\rm obs}$) at an observatory. Paper~I (Hobbs, Edwards \& Manchester 2006)\nocite{hem06} and Paper~II (Edwards, Hobbs \& Manchester 2006)\nocite{ehm06} of this series detail how the new pulsar timing package, \textsc{tempo2}\footnote{The \textsc{tempo2} software and documentation are available from our website \url{http://www.atnf.csiro.au/research/pulsar/tempo2}.}, is used to convert $t_a^{\rm obs}$ to the proper time of emission, $t_e^{\rm psr}$, as
\begin{equation}
t_e^{\rm psr} = t_a^{\rm obs} - \Delta_\odot - \Delta_{\rm IS} - \Delta_{\rm B}.
\end{equation}
$\Delta_\odot$ is the transformation required to convert the site arrival times to the solar system barycentre, $\Delta_{\rm IS}$ is the excess propagation delay due to the interstellar medium and $\Delta_{\rm B}$ is the transformation to the pulsar frame for binary pulsars. \textsc{Tempo2} compares the derived time of emission with a pulsar model to form ``timing residuals'', which are equivalently the deviations between the observed TOAs and the model predictions.  For a perfect pulsar model, random receiver noise and no other systematic effects, these timing residuals will have a mean of zero and be uncorrelated, corresponding to a flat, or ``white'', spectrum.  Since \textsc{tempo2} does not include GW sources in the timing model, the existence of any such sources will induce a signal in the timing residuals. The aim of this paper is to describe how this signal can be simulated and how such simulations aid searches for GW signals within our existing data-sets.

Since the intrinsic pulsar pulse period, spin-down, orbital motion and various astrometric parameters are a priori unknown, they must be determined from the pulsar timing data.  In common with other pulsar timing analysis programs, \textsc{tempo2} uses initial estimates of the pulsar parameters to obtain ``pre-fit'' timing residuals and then uses a least-squares fitting procedure to fit an analytical model to obtain improved pulsar parameter estimations and ``post-fit'' timing residuals (full details are given in Paper~I). The net outcome of this process is that a polynomial and various spectral components are removed from the post-fit timing residuals.  Any GW signal with a period larger than the time-span of the data is largely absorbed by the removal of the low-order polynomial terms.  Hence, pulsar timing experiments are only sensitive to GW signals with periods less than, or equal to, the time-span of the data (typically years), corresponding to frequencies in the range 1-30\,nHz.

The three basic types of GW sources that have been discussed in the literature are (1) continuous wave sources (Peters 1964)\nocite{pet64}, (2) burst sources (e.g. Thorne \& Braginskii 1976\nocite{tb76}, Damour \& Vilenkin 2001\nocite{dv01}, Kocsis et al. 2006\nocite{kgm06} and Enoki \& Nagashima 2007\nocite{en07}) and (3) stochastic backgrounds (e.g. Jaffe \& Backer 2003\nocite{jb03}, Wyithe \& Loeb 2003, Maggiore 2000\nocite{mag00}).  The GW strain spectrum for a stochastic background is thought to be a power-law in the GW frequency, $f_g$, as
\begin{equation}\label{eqn:hc}
h_c(f_g) = A_g \left(\frac{f_g}{{f_{\rm 1 yr}}}\right)^\alpha,
\end{equation}
where $f_{\rm 1yr} = 1/{\rm 1 yr}$ and $A_g$ is dimensionless. 
For a background generated by supermassive binary black holes, $\alpha = -2/3$ and $A_g \sim 10^{-15}$ (Jaffe \& Backer 2003, Wyithe \& Loeb 2003). Standard models of inflation (e.g., Turner 1997; Boyle \& Buonanno 2007)\nocite{tur97} produce GW backgrounds with amplitudes well below detectable limits with current experiments ($A_g \sim 10^{-18}$), but some non-standard models (e.g., Grishchuk 2005\nocite{gri05}) have $\alpha \sim -1$ and $A_g \sim 10^{-15}$.  Cosmic string cusps are also expected to produce a GW background with $\alpha = -7/6$ and $A_g$ can become as large as $10^{-14}$ \cite{dv01,cbs96}.

%Various upper-bounds on the GW background amplitude, $A_g$, have been
%described in the literature (see e.g., Kaspi et
%al. 1994\nocite{ktr94}, Thorsett \& Dewey 1996\nocite{td96}, McHugh
%et al. 1996\nocite{mzvl96}) and limits on the existence of single GW
%sources have also been attempted (Lommen \& Backer 2001\nocite{lb01},
%Jenet et al. 2004\nocite{jllw04}).  However, these analyses did not
%take into account all of the factors that affect real observations of
%pulsars, e.g. irregular sampling, non-`white' noise due to
%calibration problems and intrinsic pulsar timing noise, the fitting
%of the pulsar rotational, astrometric and orbital parameters and
%inaccuracies in the terrestrial time-standard or in the planetary
%ephemeris.  

Determining a rigorous limit on $A_g$ is not trivial as real pulsar data-sets have irregular sampling, non-white noise due to instrumental problems, intrinsic pulsar timing noise, astrometric and orbital parameter fitting, and inaccuracies in the terrestrial time-standard or in the planetary ephemeris.  Jenet et al. (2006)\nocite{jhv+07} recently described how simulating GW signals within \textsc{tempo2} allows rigorous limits to be placed on $A_g$, which take into account the majority of the issues affecting real pulsar observations. The use of \textsc{tempo2} and the methods employed were only outlined in the Jenet el al. paper; full details are provided here. Unfortunately, the Jenet et al. technique can only be applied to timing residuals that have a white spectrum. We have recently developed a new technique that makes no assumption on the spectrum of the timing residuals. This recent work will be presented in a forthcoming paper.

In \S2 we provide the mathematical framework that allows \textsc{tempo2} to simulate GW sources. This is divided into sections considering the timing effects induced by non-evolving GW sources (\S2.1) and evolving sources (\S2.2).  In \S3 we demonstrate applications of this mathematical framework within \textsc{tempo2}.

\section{Simulating the effect of GW sources on pulsar TOAs}\label{sec:sat}

The equations presented in this paper describe how the induced timing residuals for a given pulsar due to a GW signal can be calculated. However, this is not sufficient for our purposes. We must be able to simulate the effects of a GW on the actual pulse TOAs, because the process of fitting a timing model to obtain the residuals will modify the effects of a GW.  Numerous methods exist within \textsc{tempo2} to simulate such TOAs. These methods are all based on the following iterative procedure.  First, a set of observation dates and times are defined by the user.  Second, the entire \textsc{tempo2} timing procedure (as described in Paper~II) is carried out in order to obtain pre-fit timing residuals. This procedure uses a user-specified timing model defining the pulsar being simulated and assumes that the dates and times described above represent pulse TOAs.  Third, these pre-fit timing residuals, which really describe the timing model, are subtracted from the original arrival times. The goal is to obtain arrival times which, when fitted with a timing model, give zero residuals.  However, because of various non-linear operations in the modelling and fitting process, this procedure must be iterated until the resulting pre-fit residuals are adequately close to zero for the simulation being planned.  These TOAs can subsequently be modified by the addition of white Gaussian noise, a model of the pulsar timing noise and/or the GW signal.  The final TOAs are stored as if they were actual pulsar observations and can be processed using standard fitting and analysis routines.

\textsc{tempo2} employs two different techniques to simulate the effect of GWs on pulsar timing residuals. The first technique is used for constant frequency (i.e. ``non-evolving'') sources. The second is used for simulating the effects of binary systems that are evolving. Since the latter technique is computationally expensive, the former is used to calculate the effects of a stochastic background of GWs.  The basic implementation into \textsc{tempo2} is described below.  We also provide detailed derivations of all the main equations in the Appendix.

\begin{figure}
\begin{center}\includegraphics[width=8cm]{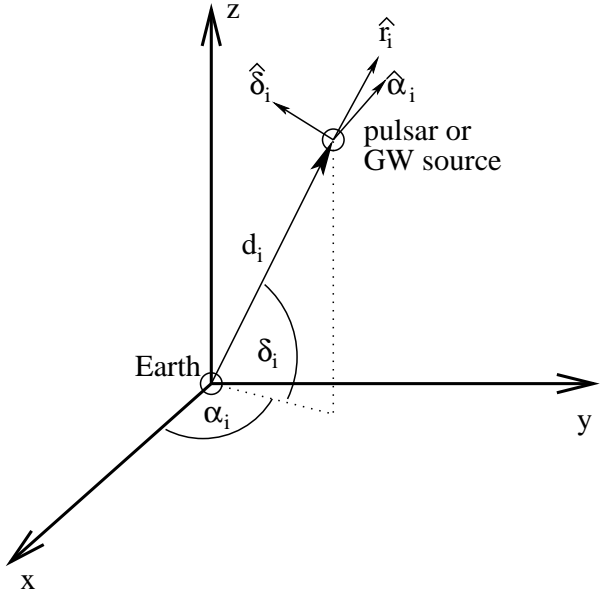}\end{center}
\caption{Configuration of the coordinate system used throughout this paper. Note that $\hat{\alpha}\times\hat{\delta} = \hat{r}$.}\label{fg:vectors}
\end{figure}

\subsection{Non-evolving GW sources}

The majority of the GW sources that may be detectable by pulsar timing are expected to evolve over timescales much longer than the typical observation time. Hence, the non-evolving algorithm used in \textsc{tempo2} can be used in most GW simulations.

The non-evolving GW simulation algorithm in \textsc{tempo2} has been defined so that the user can input pulsar and GW source positions in an equatorial coordinate system.  Figure~\ref{fg:vectors} represents the position, with respect to the Earth, of either a pulsar (with unit position vector $\hat{\bf r}_p$ and distance $d_p$) or a GW source ($\hat{\bf r}_g$, $d_g$).  The sources are specified by their right
ascension and declination ($\alpha_i$, $\delta_i$). 

\textsc{Tempo2} assumes a globally flat coordinate system with three
spatial coordinates ($x$,$y$,$z$) and one temporal coordinate $t$. GWs
are treated as a tensor field in this background space-time. A single
plane GW takes the form
\begin{equation}
h_{lm} = \mbox{Re}\left[  A_{lm} e^{i(\vec{k}_g \cdot \vec{x} - \omega_g t)}\right]
\end{equation}
where the indices $l$ and $m$ range from 1-3 corresponding to the three spatial coordinates. $A_{lm}$ is a constant tensor amplitude, $\vec{k}_g$ is the three-dimensional GW vector and $\omega_g$ is the GW angular frequency. A GW signal causes fluctuations in the observed pulsar's observed spin frequency $\delta f/f$. The induced pulsar timing residuals are given by the integral of this quantity over time. The timing residuals induced by a GW of the above form are given by
\begin{eqnarray}
R(t) = -\frac{1}{2} \mbox{Re}\left[\frac{\hat{r}^{l}_{p}A_{lm }\hat{r}^{m}_{p}}{\omega_g} \left(e^{-i \omega_{g} t} -1 \right)
  \left(\frac{1 - e^{i \omega_{g} d_p\zeta}}{\zeta}\right)\right], \label{gw_induced_r}
\end{eqnarray} 
where  $\zeta = 1-\cos\theta$ and $\theta$ is the angle between the direction of the pulsar and the direction of the GW source (see Appendix).

In \textsc{tempo2}, the GW tensor amplitude is specified in the $(\hat{r}_g, \hat{\alpha}_g,\hat{\delta}_g)$ coordinate system where the GW is propagating along the $-\hat{r}_g$ direction. GWs consistent with Einstein's equations have two independent degrees of freedom, which we label as $A_+$ and $A_\times$. Written in terms of these values, the tensor amplitude takes the form:

\begin{eqnarray}
A_{lm} = \left(\begin{tabular}{lll}
0 & 0 & 0\\
0 & $A_+$ & $A_\times$\\
0 & $A_\times$ & $-A_+$
\label{tensor_amp}
\end{tabular}\right)
\end{eqnarray}
Since \textsc{tempo2} allows one to arbitrarily specify the entire tensor amplitude, one can generate GWs consistent with any general metric theory. Once the GW amplitude is specified in the ($\hat{r}_g,\hat{\alpha}_g,\hat{\delta}_g$) coordinate system, $\hat{r}_p^l A_{lm} \hat{r}_p^m$ is evaluated by transforming both $\hat{r}_p$ and $A_{lm}$ into the global $(x,y,z)$ coordinate system. This scalar quantity is then used in equation \ref{gw_induced_r} to calculate the induced pulsar timing residuals for the given pulsar. In the remainder of this section, we will discuss how the above general framework is used to simulate GWs from a single, non-evolving, binary system as well as from a stochastic background of GWs.

\subsubsection{GWs from supermassive black-hole binary systems}

Supermassive black-hole binary systems in the cores of galaxies are expected to be sources of detectable GWs. For long-period binary systems, the time it takes for the orbital period to evolve under the action of GW emission ($\sim 10^4$ years for a system with chirp mass $M_c = 10^9$ solar masses\footnote{The chirp mass is defined as $M_c = (m_1+m_2)\left(\frac{m_1m_2}{(m_1+m_2)^2}\right)^{3/5}$ where $m_1$ and $m_2$ are the masses of the binary companions.} and a three-year orbital period) is much longer than any reasonable observation time. Hence, the binary system may be treated as non-evolving. In general, the GWs emitted by a binary system will be elliptically polarised (Blanchet et al. 1996)\nocite{blww96}. Since the tensor amplitude is a complex quantity, the effects of such GWs can be calculated using the framework described above.
 
In the current \textsc{tempo2} implementation, only systems with zero eccentricity are considered for the non-evolving case. This is a valid assumption since binary systems tend towards zero eccentricity much faster then the decay timescale (Peters 1964)\nocite{pet64}.

Following Wahlquist (1987)\nocite{wah87}, \textsc{tempo2} models GWs emitted from a binary system by setting $A_+$ and $A_\times$ as follows:
\begin{eqnarray}
A_+ = -A_g e^{-i \theta_n} \left[(3 + \cos\theta_i)\cos(2\phi) +  i 4 \cos(\theta_i)\sin(2 \phi)\right]\\
A_\times = -A_g e^{-i \theta_n} \left[(3 + \cos\theta_i)\sin(2\phi) - i 4 \cos(\theta_i)\cos(2 \phi)\right]
 \end{eqnarray}
where
\begin{equation}
A_g = \frac{M_c^{5/3} \omega_o^{2/3}}{d_g},
\end{equation}
$\theta_i$ is the orbital inclination angle, $\phi$ is the orientation of the line of nodes, $\theta_n$ is the orbital phase angle at the line of nodes, $M_c$ is the binary chirp mass, $\omega_o$ is the orbital frequency and $d_g$ is the distance to the source. Note that the GW angular frequency, $\omega_g =  2 \omega_o$.

\subsubsection{A stochastic background of GWs}

It is possible within \textsc{tempo2} to specify a large number of individual GW sources, each with different properties. A stochastic background of GWs is simulated by randomly specifying the source directions and tensor amplitudes of the GWs generated by these sources. Such a background is described by its characteristic strain spectrum, $h_c(f)$ (equation~\ref{eqn:hc}). In order to simulate such a background, probability distributions for the GW parameters are defined as follows. The source directions are chosen uniformly on the celestial sphere so that the respective probability distribution functions are given by:
\begin{eqnarray}
P(\sin \delta) &=& \frac{1}{2},\\
P(\alpha) &=& \frac{1}{2 \pi}.
\end{eqnarray}
The GW frequencies are chosen to be uniformly distributed in $\log \omega_g$:
\begin{equation}
P(\omega_g) = \left\{ \begin{array}{cc} \frac{1}{\omega_g} \frac{1}{\log(\frac{\omega_h}{\omega_l})} & \omega_l \leq \omega_g \leq \omega_h \\ 0 & \mbox{otherwise} \end{array} \right.
\end{equation}
where $\omega_l$ and $\omega_h$ can be defined by the user, but default to $\omega_h=2\pi/({\rm 1 d})$ and $\omega_l = 2\pi\frac{0.01}{T}$ where $T$ is the time-span of the observations.

$A_+$ and $A_\times$, the parameters used to determine the tensor amplitude (see equation \ref{tensor_amp}), are treated as real numbers (i.e. the imaginary parts are set to zero) that are normally distributed with zero mean and rms given by
\begin{equation}
\sigma_A(f) = \sqrt{\frac{\log(\omega_h/\omega_l)}{N}} h_c(f)
\label{sigmaA}
\end{equation}
where $N$ is the number of individual plane waves used to generate the background.

% \begin{figure}
 % \includegraphics[width=6cm,angle=-90]{J0437-4715_quad.ps}
 % \caption{The timing residuals induced due to a simulated GW
 % background of coalescing supermassive binary black hole systems for
 % PSRs~J0437$-$4715.}\label{fg:bkgrd}
 % \end{figure}

Given the above choice of distributions, the simulated background will be isotropic, unpolarised and have a Gaussian amplitude distribution with characteristic strain $h_c(f)$. For the existing simulations within \textsc{tempo2}, $h_c(f)$ is taken to be of the form given by equation~\ref{eqn:hc}. The spectral index, $\alpha$, and the amplitude, $A_g$, depend on the physical processes generating the background and may be specified by the user. 

\subsection{Evolving sources}

In general, a binary system will evolve under the action of GW emission and can have non-zero eccentricity. Although it is possible to use the above framework to model such a source, it is not very convenient. A separate module (the \textsc{GWevolve} plug-in; see below) has been developed in \textsc{tempo2} to deal with this case.  Full details of the equations integrated numerically within \textsc{tempo2} were provided by Jenet et al. (2004)\nocite{jllw04} and therefore are not reproduced here.  As shown in \S\ref{sec:3c} the user inputs the initial eccentricity and orbital periods to obtain the resulting timing residuals generated using the specified geometry of the orbit.

Unfortunately, solving the differential equations numerically is
computationally expensive.  Hence, this module is currently only used
to simulate well-defined, individual evolving sources.

\section{Applications within \textsc{tempo2}}

As described in Paper~I, the \textsc{tempo2} software is based around `plug-ins' that add to the functionality of the package. The mathematical framework described above allows for the development of new plug-ins to simulate, study, and detect GW signals.  A listing of the current plug-ins available for GW research is presented in \S\ref{sec:plugins}.  These existing plug-ins are divided into those 1) simulating the induced timing residuals due to single GW sources or from a stochastic GW background, 2) producing an upper bound on the amplitude of any GW background, 3) determining the sensitivity of a given set of pulsar timing residuals to single GW sources and 4) for inspecting the resulting timing residuals. In this section, we demonstrate these plug-ins. 

\begin{figure}
  \includegraphics[width=7cm,angle=-90]{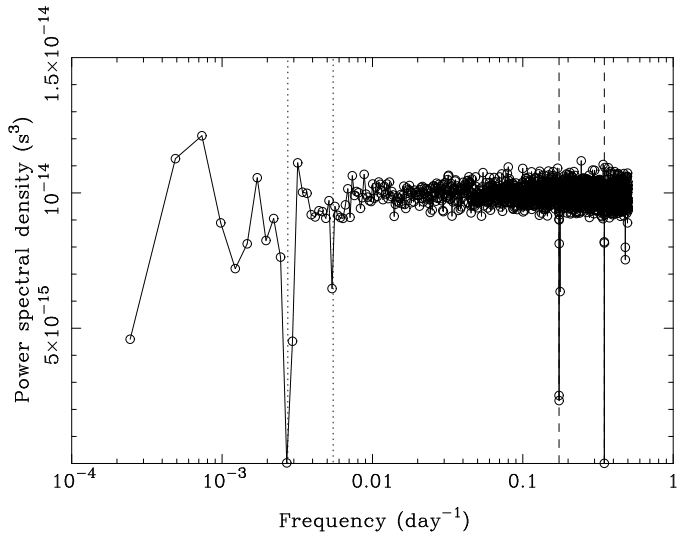}
  \caption{Average power spectrum obtained from 1000 realisations of white timing residuals and fitted using the PSR~J0437$-$4715 timing model.  The vertical dotted lines correspond to periodicities of 1\,yr and 0.5\,yr respectively.  The dashed lines correspond to the orbital period of 5.7\,d and twice the orbital period respectively.}\label{fg:absorption}
 \end{figure}

It is important that \textsc{tempo2} is used for fitting the pulsar's astrometric, pulse and, if applicable, orbital parameters when studying the induced timing residuals due to GW signals as such fits reduce the detection sensitivity at various characteristic frequencies.  In Figure~\ref{fg:absorption} we show the average power spectrum obtained after fitting a standard pulsar timing model to a white-noise, daily sampled data-set with an rms timing residual of 100\,ns. The absorption features are due to the removal of power by fitting for the astrometric, rotational and orbital parameters.  It is difficult to obtain a straight-forward description of these spectral features as they depend on the details of the fitting procedure and on the sampling of the data. However, in the \textsc{tempo2}  routines for simulating and studying GW signals that are described below, detailed analytic descriptions of such spectral features are not required; the simulations take all such effects into account.

\subsection{Stochastic GW background}

  \begin{figure}
  \includegraphics[width=7cm,angle=-90]{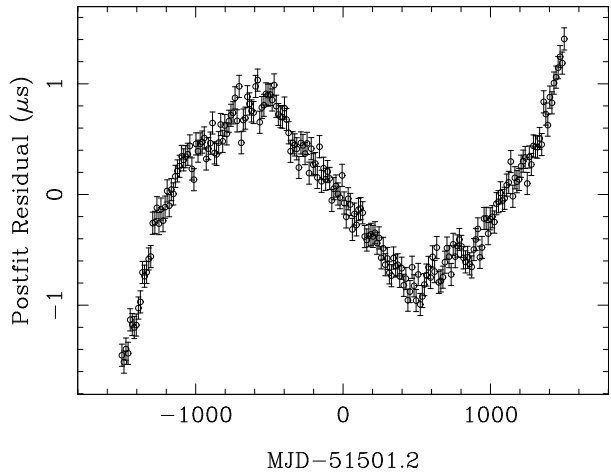}
  \caption{Example timing residuals induced by a stochastic GW background (with $A_g=10^{-14}$) after fitting for the pulsar's (PSR~B1937+21) pulse period and its first derivative.  The error bars correspond to 100\,ns of additional white, Gaussian noise.}\label{fg:exBkgrd}
 \end{figure}

GW backgrounds can be simulated using the \textsc{GWbkgrd} plug-in.  The power-law spectrum of $h_c(f)$ leads to a power-law spectrum for the pulsar timing residuals with spectral exponent $\alpha_{\rm res} = 2\alpha - 3$ (see equation~\ref{eqn:alpha}). Hence, for a background generated by supermassive black holes where $\alpha = -2/3$, the induced timing residuals will have a much steeper red-noise spectrum with spectral exponent $\alpha_{\rm res} = -13/3$. Example timing residuals (simulated every two weeks for 3000\,d) are shown in Figure~\ref{fg:exBkgrd} for PSR~B1937$+$21 after fitting for the pulsar's pulse frequency and its first derivative. 

\begin{figure}
  \includegraphics[width=8cm]{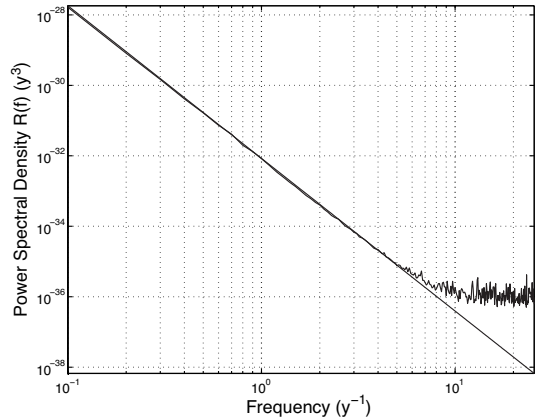}
  \caption{The average spectrum of 1000 GW background realisations for $A_g = 10^{-15}$ and $\alpha = -2/3$ for 512 weekly-spaced simulated observations. The solid, diagonal line is the theoretical spectral density.}\label{fg:gwspec}
 \end{figure}

In Figure~\ref{fg:gwspec} we show the power spectrum\footnote{Analysis of such steep 'red' spectra is challenging because of the irregular sampling of the observations and spectral leakage from the low-frequency components. In this case the sampling is regular and leakage was eliminated by prewhitening the time series with a second difference filter and postdarkening the spectrum with the inverse of the transfer function of the second difference filter. We have found that most observations can be handled with combinations of interpolation and prewhitening. These techniques are being integrated into \textsc{tempo2} and will be discussed elsewhere.} of 512 weekly sampled simulated residuals induced by a GW stochastic background with $A_g = 10^{-15}$ and $\alpha = -2/3$. The simulation was repeated 1000 times and the average power spectrum is shown, with the theoretical spectrum $(A_g^2/12\pi^2) f^{-13/3}$\,yr$^3$ drawn as a solid line. In each simulation 10000 plane GWs were summed as discussed in \S2.1.2.  On this scale the average power spectrum can barely be distinguished from the theoretical line, except at high frequencies.  The apparent high frequency noise corresponds to an rms of 0.2\,ns and occurs due to rounding errors in the pulsar timing model computations.   Since \textsc{tempo2} has been designed to maintain 1\,ns precision, and our best observations currently have an rms residual of 50\,ns, this white noise is negligible.

\subsubsection{Producing an upper-bound on the background}

Many techniques have been described in the literature for determining an upper bound on $A_g$. The earliest work (e.g. Kaspi et al. 1994\nocite{ktr94}, McHugh et al. 1996\nocite{mzvl96}) was based on analysing the measured post-fit timing residuals.  More recently, Jenet et al. (2006)\nocite{jhv+06} used the \textsc{tempo2} simulations of GW backgrounds that are described in this paper to produce an upper bound on $A_g$ for various values of $\alpha$.  This technique has limitations.  Notably it requires that the observed timing residuals are `white' (defined as being a data-set whose power spectrum is independent of frequency, or equivalently, for which the data points have no temporal correlation).  Each data-set used in the Jenet et al. (2006) work was tested by 1) forming power spectra (constructed using a Lomb-Scargle periodogram and using Gram-Schmidt orthonormal polynomials) and searching for significant periodicities and 2) averaging adjacent points to confirm that the variance of the timing residual decreases with the number of points averaged.  However, the power spectrum at low frequencies is suppressed by the fitting procedure carried out by \textsc{tempo2} and therefore even though a data-set may pass the tests described above, it may not have a purely white spectrum.

For completeness, we describe here the details of the \textsc{tempo2} usage in the original Jenet et al. (2006) method, but emphasise that new techniques are currently being developed that are not restricted to white data-sets.  It is expected that an implementation of many of these new techniques (e.g. van Haasteren et al., in press, Anholm et al., in press) will use and develop the \textsc{tempo2} functionality that is described below.

In the Jenet et al. (2006) method, a statistic is first defined that is sensitive to a GW background.  Following the terminology of the original paper we define each pulsar data-set to consist of $n_p$ measured residuals, $x_p(i)$, a time tag $t_p(i)$ and an uncertainty $\sigma_p(i)$ where $i$ is the data sample index and $p$ is an index referring to a particular pulsar.  Each data-set may be unevenly sampled. Normalised time tags
\begin{eqnarray}
\tau_p(i) = 2 (t_p(i)-t^p_{\rm min})/(t^p_{\rm  max}-t^p_{\rm min}) - 1
\end{eqnarray}
are defined where $t^p_{\rm min}$ is the earliest time and $t^p_{\rm max}$ the time of the most recent observation for pulsar $p$. Hence, $\tau_p(i)$ runs from $-1$ to $1$. These $\tau_p(i)$ values are used in a weighted Gram-Schmidt orthogonalisation procedure to determine a set of orthonormal polynomials, $j_p^l(i)$, defined from
  \begin{eqnarray}
   \sum_{i=0}^{n_p-1} \frac{j_p^l(i) j_p^k(i)}{\sigma_p^2(i)} = \delta_{lk}
  \end{eqnarray}
  where $j_p^l(i)$ is the $l$'th order polynomial evaluated at  $\tau_p(i)$ and $\delta_{lk}$ is the standard Kronecker delta  function.  The following coefficients are calculated using the  orthonormal polynomials, $j_p^l(i)$, and the timing residuals,  $x_p(i)$:

\begin{equation}
  C^l_p = \sum_{i=0}^{n_p-1}\frac{j_p^l(i)x_p(i)}{\sigma_p^2(i)}.
\end{equation}
The pulsar average polynomial spectrum is given by
\begin{equation}
   P_l = \sum_p \frac{(C^l_p)^2}{v_p}
\end{equation}
where the weighted variance, $v_p$, is defined as $\frac{1}{n_p} \sum_{i=0}^{n_p-1}(x_p(i)-\bar{x}_p)^2/\sigma_p^2(i)$ and $\bar{x}$ is the mean of $x$. For a stochastic background dominated by low-frequency noise, $P_l$ will be large for low values of $l$. Hence, $\Upsilon = \sum_{l=0}^{l=n}P_l$ is used as a statistic to detect the background. The upper limit, $n$, can be selected by the user, but $n=7$ was used throughout the Jenet et al. (2006) paper.  

The background will be ``detected'' if $\Upsilon > \Upsilon_0$ where $\Upsilon_0$ is set so that the false-alarm probability is given by ${\cal P}_f$. By default ${\cal P}_f=$\,0.1\%. $\Upsilon_0$ is obtained using the following Monte-Carlo procedure. First, standard pulsar timing procedures are followed to obtain ``pre-fit'' timing residuals, $R_1^p(i)$, for each pulsar data-set.  These are subtracted from the original site-arrival-times and the procedure iterated until arrival times, $t^p_1(i)$, are obtained that are exactly predicted by each pulsar's timing model. Noise is then added back to the arrival times. Since only pulsar residuals that are consistent with ``white noise'' can be analysed by the Jenet et al. (2006) method, an independent data set with the same noise distribution as the original is obtained by adding a shuffled version of the timing residuals $R_1^p(i)$ to $t^p_1(i)$. With this new simulated set of site-arrival-times, the entire \textsc{tempo2} timing procedure is repeated in order to obtain a new set of ``post-fit'' timing residuals, $R^p_2(i)$.  The detection algorithm is subsequently applied to $R^p_2(i)$ and the output statistic $\Upsilon_j$ is recorded.  This procedure is repeated for $N_{\rm it}$ iterations where $N_{\rm it}$ is set, by default, to 10000. These $\Upsilon_j$ values are subsequently inspected to determine $\Upsilon_0$.

Finally, the upper bound on $A_g$ is determined so that the probability of detecting the background with $A_g = A_{\rm upper}$ is ${\cal P}_d$.  By default, ${\cal P}_d = 95$\%.  This upper bound is determined by adding a GW background of a given amplitude to $t_2^p(i)$.  As above, the fitting procedures are carried out to obtain ``post-fit'' timing residuals and the detection algorithm applied to obtain $\Upsilon$.  If $\Upsilon > \Upsilon_0$ then the background has been detected.  The amplitude is changed using a bracketing procedure in order to determine the amplitude $A_{\rm  upper}$ which gives a detection probability of ${\cal P}_d$.

In this technique, the actual timing residuals  are used only as a mechanism for generating instances of white noise in the simulations.  If the spectrum of the measured timing residuals is red then this technique will provide an upper bound which is too low because the shuffled observations (which will be white) will give lower detection statistics than a simulation based on the correct noise spectrum. A plug-in package, \textsc{checkWhite}, is available in \textsc{tempo2} to test the ``whiteness'' of a data-set; see \S\ref{sec:plugins}.

The default values of  $N_{\rm it} $ and $N_{\rm gw}$ have been chosen to produce a stable upper limit that has the precision necessary for current astrophysical applications.  To demonstrate this, we use the data set for PSR~J1857$+$0943 that was first described by Kaspi et al. (1994)\nocite{ktr94} and used to determine an upper bound by Jenet et al. (2006) of $A < 1.45\times 10^{-14}$ (corresponding to a bound on the energy density per unit logarithmic frequency interval of $\Omega_{\rm gw}[1/(8 {\rm yr})]h^2 < 1.3\times10^{-7}$) for $\alpha = -1$.  Multiple simulations using the same observations, but with different realisations of the GW background and with different shuffles of the data, produces a mean upper bound of $A < 1.54 \times 10^{-14}$ and standard deviation of $0.06 \times 10^{-14}$.  It should be noted that the ``whiteness'' of the residuals of PSR~J1857$+$0943 is suspect because the observed detection statistic is 2.4 times higher than the mean of the simulated detection statistics using shuffled observations.  A detection statistic would exceed this value only 3\% of time by chance, suggesting that the residuals are somewhat red.
 
\subsubsection{Detecting the background}

\begin{figure}
  \includegraphics[width=6.0cm,angle=-90]{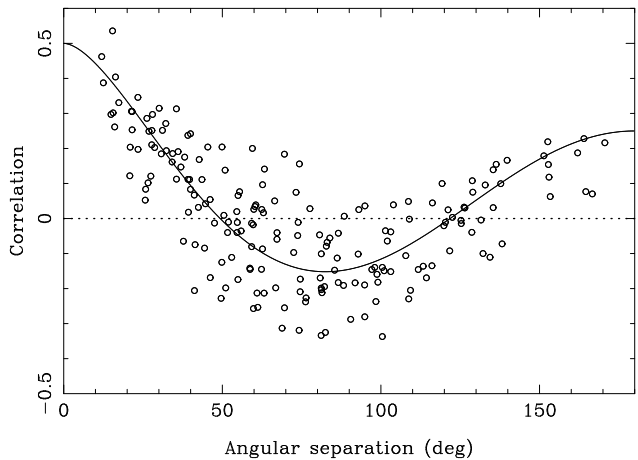}
  \caption{Pair-wise angular correlation curves for 20 simulated pulsar data-sets
  in the presence of a gravitational wave background with power-law
  index $\alpha=+3/2$ and amplitude $A_g = 0.01$.}\label{fg:corr}
 \end{figure}

Hellings \& Downs (1983)\nocite{hd83} showed that a GW background signal can be detected by searching for correlations in the timing residuals of many pulsars spread over the sky.  Within the framework of general relativity, the induced timing residuals for any isotropic, stochastic GW background are correlated with a well-defined zero-lag angular correlation function:
\begin{equation}\label{eqn:hd}
  c(\theta) = \frac{3}{2}x\ln x -\frac{x}{4}+\frac{1}{2}+\frac{1}{2}\delta(x)
\end{equation}
where $x=[1-\cos\theta]/2$ for angle $\theta$ on the sky between two pulsars. Our simulations successfully reproduce this angular correlation.  In Figure~\ref{fg:corr} we show the results of simulated timing residuals (using the \textsc{GWbkgrd} plug-in) in the presence of a GW background for the 20 PPTA millisecond pulsars (no pulsar noise is added).  For each pulsar pair, we plot the zero-lag correlation versus the angular separation of the pulsars on the sky. The solid line is the predicted functional form (equation~\ref{eqn:hd}).  In order to produce the figure using standard correlation techniques, we have selected the GW spectral exponent $\alpha = +3/2$ which corresponds to $\alpha_{\rm res} = 0$ (i.e. white timing residuals) and have simulated regularly sampled timing residuals with two weekly sampling over five years.   In general, obtaining the zero-lag correlations between pulsar pairs in the presence of red-noise in the timing residuals and uneven sampling is challenging and requires pre-whitening of the data in order to attain the maximum possible signal-to-noise ratio.  Hence, Figure~\ref{fg:corr} displays the optimal effect that could be achieved with pre-whitening techniques. These algorithms, and their implications for GW background detection, will be described in a subsequent paper.
 
\subsection{Simulating single sources and the effect of parameter fitting}\label{sec:3c}

%\begin{figure}
%  \includegraphics[width=7cm]{095DetecnCurve.eps}
% \includegraphics[width=9cm]{PLOT.eps}
 % \caption{The sensitivity of a single pulsar with 100\,ns `white' timing residuals to a sinusoidal signal at a given frequency.  The reductions in sensitivity at low frequencies and at the marked frequencies are due to fits for the pulsar's pulse period, period derivative and its astrometric and orbital parameters.  The vertical lines are at frequencies of 1/1\,yr (solid), the orbital frequency (thick-solid) and twice the orbital frequency (dashed).}\label{fg:threshold}
%\end{figure}

\begin{figure*}
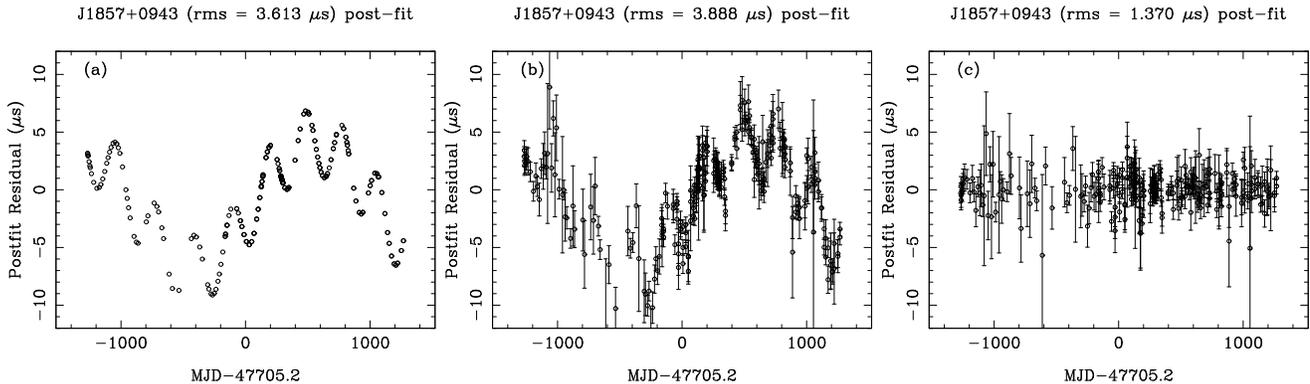

  \includegraphics[width=5cm,angle=-90]{1855_evolve1.ps}
  \includegraphics[width=5cm,angle=-90]{1855_evolve2.ps}
   \includegraphics[width=5cm,angle=-90]{1855_evolve3.ps}
  \caption{(a) Simulation of the timing residuals induced in the Kaspi et al. (1994) timing residuals for PSR~B1855$+$09 due to the postulated supermassive binary black hole system in 3C66B. (b) Simulated residuals after including the GW signal, the measured timing residuals and their uncertainties and fitting for the pulsar's spin-down, astrometric and orbital parameters. (c) The observed timing residuals.}\label{fg:3c66b}
\end{figure*}

\textsc{Tempo2} plug-ins are available to simulate both non-evolving individual GW sources (\textsc{GWsingle}) and evolving sources (\textsc{GWevolve}).  The non-evolving source simulations can easily
be shown to produce sinusoidal residuals of the correct amplitude and phase for a given GW source and pulsar position.  
%However, the spectrum of the resulting timing residuals, after standard pulsar timing fits, leads to absorption features that may affect (or entirely remove) the GW signal.  In Figure~\ref{fg:threshold}, we show the sensitivity of a single pulsar data-set to a sinusoidal signal (such as a non-evolving single GW source) obtained after fitting a standard pulsar timing model to 15\,yrs of simulated data (sampled once every two weeks) for the millisecond pulsar PSR~B1855$+$09.  The timing procedure included fits for the pulse period, its first derivative, position, parallax, proper motion and the Keplerian orbital parameters. The plot gives the amplitude of a sinusoid that would be detected at a 95\% confidence level. In the \textsc{tempo2} procedure, detailed analytic descriptions of these absorption features are not required; the simulations take all such effects into account.
In Figure~\ref{fg:3c66b} we use the \textsc{GWevolve} plug-in to reproduce the expected PSR~B1855$+$09 timing residuals for the postulated binary supermassive black-hole system in the radio galaxy 3C66B (Sudou et al. 2003)\nocite{simt03}.  As described in Jenet et al. (2004), the induced signal has a low-frequency component due to the GW signal at the pulsar and a higher frequency component due to the GW signal at Earth. Figure~\ref{fg:3c66b}a shows the pre-fit timing residuals induced by the simulated GWs. Figure~\ref{fg:3c66b}b gives a realistic representation of observed post-fit timing residuals if 3C66B did contain a binary black-hole system with the chirp mass and period given in Sudou et al. (2003). As concluded by Jenet et al. (2004), such a signal would be easily detectable, but has not been observed in actual pulsar data-sets (Figure~\ref{fg:3c66b}c). Note that the low-frequency term would be indistinguishable from the cubic variations often observed and attributed to pulsar period irregularities.

One of the best known candidates for a supermassive binary black hole system emitting GWs with frequencies detectable by pulsar timing is in the blazar OJ287, where a periodicity of $\sim 12$\,yr has been identified in optical outbursts (e.g., Sillanpaa et al. 1996)\nocite{stp+96}.  The parameters of this system are not well-defined. However, to obtain an order-of-magnitude estimate of the induced timing residuals due to the GW emission from this system, we can use the parameters originally suggested by Sillanpaa et al. (1988) and make no cosmological corrections.  They model the system with $m_1 = 2\times10^7$\,M$_\odot$, $m_2 = 5\times10^9$\,M$_\odot$, an orbital period of 9\,yr in the rest frame of the blazar and an initial eccentricity of $e = 0.7$.  The source has a red-shift of 0.306 corresponding to a distance of $\sim$1250\,Mpc.  The \textsc{GWevolve} plug-in shows that the induced timing residuals due to this system are significantly less than 1\,ns and therefore undetectable in all existing data-sets. Note that cosmological effects will only change this result by about 20\%.

\subsection{Public data sets}\label{sec:files}

Many publications which have described limits on the existence of a GW background, or on the existence of individual GW sources, have relied on publically available pulsar timing residuals
(in particular, most have used the data sets made available by Kaspi et al. 1994)\nocite{ktr94}. It is likely that the resurgence of interest in pulsar timing arrays and GW detection will lead to many more techniques being developed.  In order to aid comparison between different techniques we have made available a set of simulated pulsar timing residuals with and without the addition of a GW background.  These timing residuals, arrival time files and parameter models are available from our web-site\footnote{select the ``publically available data files'' link from http://www.atnf.csiro.au/research/pulsar/tempo2}.  The timing residuals for the simulated PPTA data are also available as an electronic supplement to this paper (see Appendix~\ref{sec:electronicTables}).

\begin{itemize}
\item{Simulated PPTA data: based on the design specifications for the PPTA project, we provide data-sets with two-weekly sampling of 20 pulsars with white, Gaussian noise giving 100\,ns rms timing residuals.  We include data-sets 1) without the addition of a GW background, 2) with a background where $A_g = 10^{-14}$ and $\alpha = -2/3$ and 3) with a background where $A_g = 10^{-15}$. The pulsar parameter files were obtained from the ATNF pulsar catalogue (Manchester et al., 2005).}

\item{Simulated global timing array data: we provide data-sets which are likely to be created by the global pulsar timing array (i.e. combining observations of both Northern and Southern hemisphere pulsars).  We simulate 30 pulsars (the 20 PPTA pulsars and 10 more Northern millisecond pulsars), with data spans ranging from 5\,yr to 12\,yr, realistic observation dates and rms timing residuals from 100\,ns to 1\,$\mu s$.  We provide two different GW background amplitudes (with $A_g = 10^{-15}$ and $A_g=10^{-14}$ respectively).}

\item{Simulated SKA data: The SKA is likely to be able to time at least 100 millisecond pulsars with rms timing residuals around 50\,ns. In order to simulate possible data-sets we select the 100 fastest recycled pulsars in the ATNF pulsar catalogue that are not associated with globular clusters. We simulate weekly sampled, white timing residuals over a data-span of 10\,yr.} 
\end{itemize}

\subsection{Plug-in packages}\label{sec:plugins}

The following plug-ins are available for \textsc{tempo2} from our website.

 \begin{itemize}
  \item{\textsc{fake}: As described in \S\ref{sec:sat}, this plug-in allows the user to simulate pulse arrival-times at an observatory that are in accordance with a specified pulsar timing model to better than 1\,ns.  This plug-in has been used in producing the publically available files that are described in \S\ref{sec:files}.}
  \item{\textsc{GWbkgrd}: This plug-in allows the user to simulate the pre- and post-fit timing residuals resulting from a specified GW background.}
  \item{\textsc{GWsingle}: Allows the user to simulate the pre- and post-fit timing residuals resulting from a non-evolving super-massive black-hole binary system at a given distance.}
  \item{\textsc{GWevolve}: This plug-in determines pulse arrival-times that have been affected by a binary source evolving due to emission of gravitational radiation. }
  \item{\textsc{GWwhiteLimit}: This plug-in implements the technique first used by Jenet et al. (2006) to place an upper-bound on the amplitude of a GW background.}
  \item{\textsc{checkWhite}: A plug-in to test the ``whiteness'' of a particular data-set. This plug-in plots various power-spectral estimates (including a Lomb-Scargle periodogram and a Gram-Schmidt orthogonal polynomial power spectrum) and calculates the statistic used in the Jenet et al. (2006) upper-bound technique for the actual timing residuals and for shuffled realisations of the timing residuals.}
  \item{\textsc{plk}: This plug-in is available with the default \textsc{tempo2} distribution.  It allows the user to view pre- and post-fit pulsar timing residuals.  The user may turn on (or off) fitting for various model parameters and re-calculate post-fit timing residuals.  Figures~\ref{fg:exBkgrd} and \ref{fg:3c66b} in this paper were obtained using this plug-in.}
 \end{itemize}

\section{Conclusions}

A major advantage of \textsc{Tempo2} over previous pulsar timing packages is that its functionality can be expanded using plug-in packages.  Numerous plug-ins have now been developed in order to simulate and analyse the effects of GW signals on pulsar timing data.  This code has already been used to place the most stringent constraints to date on the existence of a GW background (Jenet et al. 2006). It is now being used to study how a GW background could be detected, to determine the sensitivity of a given pulsar timing array to single and burst GW sources and to study the possibilities of pulsar timing array projects with future instruments such as the Square Kilometre Array telescope, with which we hope to not only detect gravitational waves, but also to study their properties in detail and the sources from which they emanate.
 
\section*{Acknowledgments} 

 The Parkes Pulsar Timing Array project is a collaboration between the ATNF, Swinburne University of Technology and the University of Texas, Brownsville, and we thank our collaborators on this project.  This research was funded in part by the National Science Foundation (grant \#0545837). GH is the recipient of an Australian Research Council QEII Fellowship (project \#DP0878388) and RNM is the recipient of an Australian Research Council Federation Fellowship (project \#FF0348478).

\bibliography{modrefs,psrrefs,crossrefs}
\bibliographystyle{mn}

\appendix
\onecolumn
\newcommand{\sinc}{\mbox{sinc}}
\section{Mathematical description of GW sources}

This Appendix shows how the expected power spectrum from a stochastic background of GWs is calculated. This power spectrum is related to the characteristic strain spectrum, $h_c$, and the
normalized power per logarithmic frequency interval, $\Omega_{gw}$. In order to establish a consistent, well-defined notation, the calculations are presented from first principles.  Note that we are using standard geometrised units where $c=1$.

\subsection{The stochastic background and its energy density}

GWs are linear perturbations to a background space-time metric. For the purpose of this paper, we will assume that the background space-time is flat. Hence, the space-time metric may be
written as:
\begin{equation}
g_{\mu\nu} = \eta_{\mu\nu} + h_{\mu\nu}
\label{metric}
\end{equation}
where
\begin{equation}
\eta_{\mu\nu} = \left(\begin{tabular}{c c c c} -1&0&0&0\\ 0&1&0&0 \\ 0&0&1&0\\ 0&0&0&1\end{tabular}\right)
\end{equation}
and $h_{\mu\nu}$ is a small perturbation. The linearised Einstein equations with $\eta_{\mu\nu}$ as the background space-time take the following form:
\begin{equation}
h^\lambda_{\sp\lambda,\mu\nu}  - h^\lambda_{\sp \mu,\lambda\nu}   -h^\lambda_{\sp\nu,\lambda\mu} + h_{\mu\nu,\lambda}^\lambda = 0.
\end{equation}

A stochastic background of GWs is made up of a sum of plane waves travelling in several directions. Hence, one can write the metric perturbation due to a stochastic background as:
\begin{equation}
h_{\mu\nu} = \mbox{Re}\left[ \sum_{j=0}^{N-1} A_{\mu\nu_j} e^{i\vec{k_j} \cdot \vec{x} - i\omega_j t}\right]
\label{bgd_metric}
\end{equation}
where $N$ is the total number of GWs and $A_{\mu\nu_j}$, $\vec{k_j}$ and $\omega_j$ are the complex amplitude, spatial wave vector and angular frequency of the $j$th GW respectively. $\omega_j$ is taken to be positive.

The stress-energy tensor for a metric perturbation is given by a 4-D volume average:

\begin{equation}
T^{gw}_{\alpha\beta} = \frac{1}{32 \pi}\frac{1}{T}\frac{1}{L^3}\int h_{\mu\nu,\alpha}h^{\mu\nu}_{\sp\sp,\beta} d^3x dt,
\end{equation}
where $L$ and $T$ are the spatial and temporal limits of integration respectively. $L$ and $T$ are taken to be several times the longest wavelength involved. Using the metric of a stochastic background, equation~\ref{bgd_metric}, the energy density, $\rho_{gw}$, takes the
form:
\begin{eqnarray}
\rho_{gw} = T^{gw}_{00} &=&  \frac{1}{32 \pi} \frac{1}{T}\frac{1}{L^3}\int \frac{1}{4}\sum_{j l}\left(-i\omega_j A_{\mu\nu_j}e^{i\vec{k_j} \cdot \vec{x} - \omega_j t} + i\omega_j A^{*}_{\mu\nu_j}e^{-i\vec{k_j} \cdot \vec{x} + \omega_j t}\right) \times \\ \nonumber & &\left(-i\omega_l A^{\mu\nu}_le^{i\vec{k_l} \cdot \vec{x} - \omega_l t} + i\omega_l A^{*\mu\nu}_le^{-i\vec{k_l} \cdot \vec{x} + \omega_l t}\right)d^3x dt.
\end{eqnarray}
As long as there are a finite number of plane GWs, or sources, in the sum, one can make the following approximation with reasonable accuracy:
\begin{equation}
\frac{1}{T}\frac{1}{L^3}\int e^{i (\vec{k}_j-\vec{k}_l)\cdot \vec{x} - (\omega_j - \omega_l)t}dt d^3x = \delta_{jl}
\end{equation}
where $\delta_{jl} = 1$ if $j=l$ and zero otherwise. Using this, the energy density becomes:
\begin{equation}
\rho^{gw} = \frac{1}{64 \pi} \sum_i \omega_i^2 A^{*}_{\mu\nu_i}A^{\mu\nu}_i.
\end{equation}A

Next, the above sum will be written in integral form using a probability density function. The amplitude of a given GW depends on $\vec{k}$ and a set of other parameters denoted as $\vec{\alpha}$. Examples of these other parameters are mass and distance. Letting $dP/d^n\alpha d^3k$ be the probability density for all the parameters on which a general GW may depend, the ensemble-averaged energy density is given by
\begin{equation}
\rho^{gw} = \frac{1}{64 \pi} \int \omega^2 A^{*}_{\mu\nu}(\vec{k},\vec{\alpha})A^{\mu\nu}(\vec{k},\vec{\alpha}) N \frac{dP}{d^n\alpha d^3 k} d^n\alpha d^3k \label{t00}.
\end{equation}

Since $d^3k = \omega^2 d\omega d\Omega$ where $d\Omega = d\cos\theta d\phi$ and $\theta$ and $\phi$ are the usual spherical coordinate angles specifying the GW propagation direction, the
energy density per unit frequency is given by
\begin{equation}
\frac{d\rho_{gw}}{d\omega} = \frac{1}{64 \pi}\int \omega^4 A^{*}_{\mu\nu}(\vec{k},\vec{\alpha})A^{\mu\nu}(\vec{k},\vec{\alpha}) N \frac{dP}{d^n\alpha d^3k} d^n\alpha d\Omega.
\end{equation}

\subsection{The stochastic background and the induced timing residuals}

The action of a  GW slightly alters the arrival times of radio pulses emitted by a radio pulsar. Equivalently, the rate of arrival of the pulses will fluctuate. Since the action of gravity does not depend on the frequency of the electromagnetic (EM) radiation, the problem of determining the change in the rate of arrival of pulses of EM radiation simplifies to the problem of finding the change in frequency of a single-frequency plane wave or photon.

The four-dimensional path of a photon is the shortest path between specified end points. The four-dimensional path is represented by $x^\mu(\lambda)$ where $\lambda$ is the so-called ``affine parameter''. At any given point along the path, the wave four-vector is determined by $k_p^\mu = dx^{\mu}/d\lambda$. Since the infinitesimal distance between two points on the four-dimensional curve is given by $\sqrt{k_p^\mu k_p^\nu g_{\mu\nu}}d\lambda$, the ``distance'' between two points along any such curve is given by:

\begin{equation}
D = \int_{\lambda_0}^{\lambda_1} \sqrt{-k_p^\mu k_p^\nu g_{\mu\nu}} d\lambda.
\end{equation}
 
Defining $L = \sqrt{-k_p^\mu k_p^\nu g_{\mu\nu}}$, the shortest path between two fixed endpoints is given by the four Lagrange equations:
\begin{equation}
\frac{d}{d\lambda}\frac{d L}{d k^{\alpha}} - \frac{d L}{d x^\alpha} = 0.
\end{equation}

Using

\begin{eqnarray}
\frac{dL}{dk^{\alpha}} &=& -\frac{1}{L}k_p^\nu g_{\alpha\nu} = -\frac{1}{L}k_{p\alpha}\\
\frac{dL}{dx^\alpha} &=& -\frac{1}{L}\frac{1}{2}k_p^\mu k_p^{\nu} g_{\mu\nu,\alpha}
\end{eqnarray}
the Lagrange equations yield:

\begin{equation}
\frac{d k_{p\alpha}}{d\lambda} = \frac{1}{2}k_p^{\mu}k_p^{\nu}g_{\mu\nu,\alpha}.
\end{equation}

In order to calculate the terms due to the action of the GW alone, we will assume that both the pulsar and the observer are at rest in the global background coordinate system. In this case, $k_{p0}$ is the frequency of the photon. $k_{p0}$ at the pulsar will be written as $\omega_e$ while at the receiver it will be denoted as $\omega_r$. Next, we will write the metric using equation \ref{metric} and let $k_p^{\mu} = \bar{k}^{\mu}_{p} + \delta k^{\mu}_0$ where $\bar{k}^{\mu}_{p}$ is the photon four-vector in the unperturbed space-time and $\delta k^{\mu}_{p}$ is the induced perturbation to the path. The equation for the perturbed photon frequency is then given
by:
\begin{equation}
\frac{d \delta k_{p0}}{d\lambda} = \frac{1}{2} \bar{k}^{\mu}\bar{k}^{\nu} h_{\mu\nu,0}.
\end{equation}
Using equation~\ref{bgd_metric} for the metric, results in
\begin{equation}
\frac{d \delta \omega}{d\lambda} = \frac{1}{2} \mbox{Re}\left[ \sum_j -i \omega k^{\mu}_{p0}k^{\nu}_{p0} A_{\mu\nu j} e^{i k_\mu x^\mu(\lambda)}\right]
\label{dk}
\end{equation}
where $x^\mu(\lambda)$ is the unperturbed photon path given by
\begin{equation}
x^\mu(\lambda) = \bar{k}_p^\mu (\lambda - \lambda_e) + x^\mu_e,
\end{equation}
$x^\mu_e$ is the location of the photon emitter (i.e. the pulsar) and $\lambda_e$ is the affine parameter of the emitter. Putting this into equation~\ref{dk} and integrating yields:
\begin{equation}
\delta \omega_r - \delta \omega_e =  \frac{1}{2} \mbox{Re}\left[\sum_j -i \omega\bar{k}^{\mu}_{p}\bar{k}^{\nu}_{p} A_{\mu\nu j} \left(\frac{e^{i k_{\mu j}x_r^\mu} - e^{i k_{\mu j} x^{\mu}_e}}{ik_{\mu j}\bar{k}_p^{\mu}}\right)\right]
\label{domega1}
\end{equation}
where $x^\mu_r$ is the location of the receiver. Using 
\begin{equation}
x^\mu_r - x^\mu_e = \bar{k}_p^\mu(\lambda_r-\lambda_e) 
\end{equation}
together with the unperturbed light travel time between the pulsar and the receiver, $D=x_r^0 - x_e^0$, it can be shown that
\begin{equation}
x^\mu_r - x^\mu_e = \frac{D}{\omega_e}\bar{k}_p^\mu.
\end{equation}
With the above, equation \ref{domega1} may be written as
\begin{equation}
\delta \omega_r - \delta \omega_e =  \frac{1}{2} \mbox{Re}\left[\sum_j -i \omega\bar{k}^{\mu}_{p}\bar{k}^{\nu}_{p} A_{\mu\nu j} e^{i k_{\mu j}x_r^\mu}\left(\frac{1 - e^{i \frac{D}{\omega_e}k_{\mu j}\bar{k}_p^{\mu}}}{ik_{\mu j}\bar{k}_p^{\mu}}\right)\right].
\end{equation}

Next, take the receiver location to be $x_r^{\mu} = (t,0,0,0)$ and write $k_{\mu j}\bar{k}^{\mu}_p$ as 
\begin{equation}
k_{\mu j}\bar{k}^{\mu}_p = -\omega_e \omega_{j}(1 - \cos\theta),
\end{equation}
where $\theta$ is the angle between the direction of the GW source and the pulsar. With the above, the fractional frequency shift may be written as
\begin{equation}
\frac{\delta \omega_r - \delta \omega_e}{\omega_e} =  \frac{1}{2} \mbox{Re}\left[\sum_j \frac{\bar{k}^{\mu}_{p}\bar{k}^{\nu}_{p}}{\omega_e^2} A_{\mu\nu j} e^{-i \omega_{j} t}\left(\frac{1 - e^{i \omega_{j} D ( 1 - \cos\theta_j)}}{1 - \cos\theta_j}\right)\right].
\end{equation}

The coordinate system has been chosen so that each $A_{\mu\nu}$ only has spatial components. Using this, the final form of the fractional frequency shift is:
\begin{equation}
\frac{\delta \omega_r - \delta \omega_e}{\omega_e} =  \frac{1}{2} \mbox{Re}\left[\sum_j \hat{k}^{l}_{p}\hat{k}^{m}_{p} A_{lm j} e^{-i \omega_{j} t}\left(\frac{1 - e^{i \omega_{j} D ( 1 - \cos\theta_j)}}{1 - \cos\theta_j}\right)\right]
\end{equation}
where $\hat{k}_p$ is the unit vector in the direction of the pulsar.

The change in the arrival time of a pulse at time $t$ is given by the integral of the fractional change in frequency of the pulse rate. Hence, the timing residuals induced by a set of plane GWs is given by

\begin{eqnarray}
R(t) &=& \int_0^{t} \frac{\delta \omega_r(t') - \delta \omega_e}{\omega_e} dt' 
     = -\frac{1}{2} \mbox{Re}\left[\sum_j i \frac{\hat{k}^{l}_{p}\hat{k}^{m}_{p} A_{lm j}}{\omega_j} \left(e^{-i \omega_{j} t} -1 \right)\left(\frac{1 - e^{i \omega_{j} D ( 1 - \cos\theta_j)}}{1 - \cos\theta_j}\right)\right].
\end{eqnarray} 

In order to simplify notation, the following definitions are made:
\begin{eqnarray}
B(t)_j &=& \frac{1}{2}\frac{i \left(e^{-i \omega_j t} -1 \right)}{\omega_j} \label{b}\\
C_j    &=& \frac{1 - e^{-i \omega_j D (1-\cos\theta_j)}}{1 - \cos\theta_j} \label{c}\\
E_j    &=& \hat{k}^l \hat{k}^m A_{lmj} \label{e}.
\end{eqnarray}

Using the above notation and explicitly taking the real part of the summand, the induced timing residuals become:
\begin{equation}
R(t) = - \frac{1}{2}\sum_j B_j(t) C_j E_j + B^*_j(t) C^*_j E^*_j.
\label{simp_resids}
\end{equation}

Next, the ensemble-averaged power spectrum of the residuals is calculated. For a given length of data, $T$, the variance of the residuals is given by
\begin{eqnarray}
\sigma^2 &=& \frac{1}{T}\int_0^{T} R^2(t) dt  - \left(\frac{1}{T}\int_0^T R(t) dt \right)^2 \label{sig2}.
%&=& \frac{1}{4} \sum_{j,k} \left(B_j(t) C_j E_j + B^*_j(t) C^*_j E^*_j \right) \times \nonumber\\
%& & \left(B_k(t) C_k E_k + B^*_k(t) C^*_k E^*_k \right)
\end{eqnarray}

In order to take the ensemble average of the above, it is assumed that no two GWs have the same $k^\mu$ and that GWs with different $k^\mu$ are not related to each other;  GWs from different regions of the sky are uncorrelated. Mathematically, the above statements are expressed as:
\begin{eqnarray}
\langle A_{l m j} A_{pq k}\rangle &=& 0 \\
\langle A^*_{l m j} A_{pq k}\rangle &=& \langle A^*_{lm j}A_{pq j}\rangle\delta_{jk}.
\end{eqnarray}

With the above, the ensemble average variance may be calculated by putting equation~\ref{simp_resids} into \ref{sig2}:
\begin{eqnarray}
\langle\sigma^2\rangle =&\frac{1}{2} \sum_{j}\left\langle|C_j|^2 |E_j|^2  
 \left(\frac{1}{T}\int_0^T |B_j(t)|^2 dt  - \left| \frac{1}{T} \int_0^T B_j(t) \right|^2\right)\right\rangle &\label{sigma2}
\end{eqnarray}
where $|x|$ is the complex amplitude of $x$. The integrals in the summand take the form:
\begin{eqnarray}
\frac{1}{T}\int_0^T |B_j(t)|^2 dt - \left| \frac{1}{T} \int_0^T B_j(t) \right|^2 &=& 
\frac{1}{4 \omega^2_j} \left[1 - \mbox{sinc}^2\left(\frac{\omega_j T}{2}\right) \right]
\end{eqnarray}
where sinc$(x) = \sin(x)/x$. Using the same technique to derive equation \ref{t00}, equation \ref{sigma2} may be written as:
\begin{eqnarray}
\langle\sigma^2\rangle &=& \frac{1}{4}\int \frac{1}{\omega^2}\left[1 - \mbox{sinc}^2\left(\frac{\omega T}{2}\right)\right] |\hat{k}^l \hat{k}^m A_{lm}(\vec{k},\vec{\alpha})|^2 
 \left(\frac{1 - \cos[\omega D (1 - \cos\theta)]}{(1 - \cos\theta)^2}\right) 
  N \frac{dP}{d^n\alpha d^3 k} d^n\alpha d^3k d\Omega.
 \end{eqnarray}

The fact that $k^2 = \omega^2$ implies $d^3k = \omega^2 d\omega$. Hence, the above equation tells us that the power spectrum of the residuals is given by
\begin{eqnarray}
\frac{d\sigma^2}{d\omega} &=& \frac{1}{4}\int \left[1-\sinc^2\left(\frac{\omega T}{2}\right)\right]|\hat{k}^l \hat{k}^m A_{lm}(\vec{k},\vec{\alpha})|^2 \
 \frac{1 - \cos[\omega D (1 - \cos\theta)]}{(1 - \cos\theta)^2}
 N \frac{dP}{d^n\alpha d^3 k} d^n\alpha d\Omega. \label{sigma22}
 \end{eqnarray}

\subsection{An isotropic, unpolarised,  GW background}

Until now, the derived expressions for both the energy density of a GW
background and the induced pulsar timing residuals have allowed for an
arbitrary directional dependence. Here, the calculations will be
simplified for the case of an isotropic background and the power
spectrum of the induced timing residuals will be calculated in terms
of the normalised energy density per unit logarithmic frequency interval,
$\Omega_{gw}(f)$. In this case, $dP/d^n{\alpha} d^3k$ does not depend
on the direction of the GW. The energy density per unit frequency may be written as:
\begin{equation}
\frac{d\rho_{gw}}{d\omega} = \frac{1}{16}\int \omega^4 A^{*}_{\mu\nu}(\omega,\vec{\alpha})A^{\mu\nu}(\omega,\vec{\alpha}) N \frac{dP}{d^n\alpha d^3k} d^n\alpha.
\end{equation}

$A^{*}_{\mu\nu}A^{\mu\nu}$ can be expressed in terms of the amplitudes of the two independent GW modes $A_{+}$ and $A_{\times}$:
\begin{equation}
A^{*}_{\mu\nu}A^{\mu\nu} = 2 |A_{+}|^2 + 2 |A_{\times}|^2.
\end{equation}

Since the GW background is taken to be unpolarised, $\langle|A_{+}|^2\rangle = \langle|A_{\times}|^2\rangle$. Hence
\begin{equation}
\langle A^{*}_{\mu\nu}A^{\mu\nu}\rangle = 4 |A_{+}|^2
\end{equation}
and the energy density per unit frequency may be written as
\begin{equation}
\frac{d\rho_{gw}}{d\omega} = \frac{1}{4} \omega^4 \int |A_{+}(\omega,\vec{\alpha})|^2 N \frac{dP}{d^n\alpha d^3k} d^n\alpha. \label{drhodomega}
\end{equation}

In order to calculate the induced residual power spectrum for an isotropic background, equation~\ref{sigma22} will be expressed in a standard spherical coordinate system with the pulsar located along the
$z$-axis. As before, $\theta$ represents the angle between the pulsar and the direction of the GW as well as the standard spherical coordinate polar angle. $\hat{r}$ is the unit vector pointing in the direction of the source. $\hat{\theta}$ and $\hat{\phi}$ are unit vectors pointing in the direction of increasing $\theta$ and $\phi$, respectively. These unit vectors, which depend on $\theta$ and $\phi$, make up a local right-handed coordinate system with $\hat{\theta} \times \hat{\phi} = \hat{r}$. Each $A_{ij}$ can be written in terms of the $\hat{r},\hat{\theta},\hat{\phi}$ coordinate system:
\begin{eqnarray}
A_{rr} &=& 0 \\
A_{\theta \theta} &=& -A_{\phi \phi} = A_{+}\\
A_{\theta \phi}  &=& A_{\phi\theta} = A_{\times}
\end{eqnarray}
with all other components equal to zero. Using the above, one finds that
\begin{equation}
|\hat{k}^l \hat{k}^m A_{lm}(\vec{k},\vec{\alpha})|^2= \sin^4(\theta)|A_{+}|^2.
\end{equation}
Since the pulsar lies in the $\hat{z}$ direction, $\hat{\phi} \cdot \hat{z} =0$, and $\hat{\theta} \cdot \hat{z} = -\sin\theta$.  The induced timing residuals therefore become:
\begin{eqnarray}
\frac{d\sigma^2}{d\omega} &=& \frac{1}{4}\left[1-\sinc^2\left(\frac{\omega T}{2}\right)\right] \int |A_{+}|^2  N \frac{dP}{d^n\alpha d^3 k} d^n\alpha 
\int \sin^4\theta \frac{1 - \cos[\omega D (1 - \cos\theta)]}{(1 - \cos\theta)^2} d\Omega. \label{dsigmadf}
\end{eqnarray}
Since the background is assumed to be isotropic, neither $dP/d^n\alpha d^3k$ nor $|A_{+}|^2$ depend on direction, hence they are taken outside the $d\Omega$ integral. The integration over solid angle is given by:
\begin{eqnarray}
\int_{-1}^{1} \sin^4\theta\frac{1 - \cos[\omega D (1- \cos\theta)]}{(1-\cos\theta)^2} d\Omega &=& 
\frac{16 \pi}{3} - \frac{8 \pi}{(\omega D)^2} + \frac{4 \pi \sin(2 \omega D)}{(\omega D)^3}.
\end{eqnarray}

At this point, the short wavelength approximation will be made (i.e. $\omega D >> 1$) so that the last two terms in the above are negligible. The power spectrum of the induced timing residuals can now be written as:
\begin{equation}
\frac{d\sigma^2}{d\omega} = \frac{4 \pi}{3} \left[1-\sinc^2\left(\frac{\omega T}{2}\right)\right] \int |A_{+}|^2  N \frac{dP}{d^n\alpha d^3 k} d^n\alpha. 
\label{dsdomega}
\end{equation}

Comparing this to equation \ref{drhodomega}, the power spectrum of the induced residuals may be written in terms of the energy density per unit frequency:
\begin{equation}
\frac{d\sigma^2}{d\omega} = \frac{16\pi}{3} \frac{1}{\omega^4}\frac{d\rho_{gw}}{d\omega}\left[1-\sinc^2\left(\frac{\omega T}{2}\right)\right].
\end{equation}

In terms of frequency ($f = \omega/2\pi$), this becomes:
\begin{equation}
\frac{d\sigma^2}{df} = \frac{1}{3\pi^3} \frac{1}{f^4}\frac{d\rho_{gw}}{d f}\left[1-\sinc^2\left(\frac{2 \pi f T}{2}\right)\right].
\end{equation}

Typically, the energy density spectrum is written per unit logarithmic frequency interval as:
\begin{equation}
 \frac{d\sigma^2}{df} = \frac{1}{3\pi^3} \frac{1}{f^5}\frac{d\rho_{gw}}{d \log(f)}\left[1-\sinc^2\left(\frac{2 \pi f T}{2}\right)\right].
\end{equation}

In terms of $\Omega_{gw}(f) = \frac{1}{\rho_c}\frac{d\rho_{gw}}{d\log(f)}$, one has
\begin{equation}
 \frac{d\sigma^2}{df} = \frac{H_0^2}{8 \pi^4} \frac{1}{f^5}\Omega_{gw}(f)\left[1-\sinc^2\left(\frac{2 \pi f T}{2}\right)\right]
\label{d2dfvOmega}
\end{equation}
where $\rho_c = 3 H_0^2/ 8 \pi$, and $H_0$ is the Hubble constant.

\subsection{Timing residuals and the characteristic strain spectrum}

Several investigators use the ``one-sided'' strain spectrum,
$S_h(f)$, of the GW background or the characteristic
strain spectrum $h_c(f)$. These quantities are defined as:
\begin{eqnarray}
\int_0^\infty S_h(f) df &=& \frac{1}{2} \langle h_{\mu\nu}(t)h^{\mu\nu}(t)\rangle \label{Shf}\\
h_c(f) &=& \sqrt{f S_h(f)}.
\end{eqnarray}

Using the same techniques employed above to calculate $d\rho_{gw}/df$,
one finds that
\begin{eqnarray}
\langle h_{\mu\nu}(t)h^{\mu\nu}(t)\rangle &=& \sum_j \frac{1}{2} A^{*}_{\mu\nu_j}A^{\mu\nu_j} \nonumber \\
&=& \frac{1}{2} \int A^{*}_{\mu\nu}(\vec{\alpha},\vec{k})A^{\mu\nu}(\vec{\alpha},\vec{k}) N \frac{dP}{d^n\alpha d^3 k} \omega^2 d\omega d\Omega d^n\alpha \nonumber\\
&=& 64 \pi^4 \int f^2 |A_{+}|^2 N \frac{dP}{d^n\alpha d^3k} d^n\alpha df \label{hh}
\end{eqnarray}
where the last equality holds for the case of an isotropic, unpolarised background. Using the above and the definition of $S_h(f)$, one finds that
\begin{equation}
S_h(f) = 32 \pi^4 f^2 \int |A_{+}|^2 N \frac{dP}{d^n\alpha d^3k} d^n\alpha.  
\end{equation}

Using this together with equation \ref{dsdomega}, the power spectrum of the residuals is given by:
\begin{equation}
\frac{d\sigma^2}{df} = \frac{1}{12 \pi^2}\frac{1}{f^2}S_h(f)\left[1-\sinc^2\left(\frac{2 \pi f T}{2}\right)\right] = \frac{1}{12 \pi^2}\frac{1}{f^3}h_c(f)^2\left[1-\sinc^2\left(\frac{2 \pi f T}{2}\right)\right].
\end{equation}
For the case of a power-law characteristic strain spectrum as given by equation \ref{eqn:hc}, the power spectrum of the residuals may be written as:
\begin{equation}
\frac{d\sigma^2}{df} = \frac{1}{12 \pi^2} \left(\frac{f}{f_{\rm 1yr}}\right)^{2 \alpha -3}\frac{A_g^2}{f_{\rm 1yr}^3}\left[1-\sinc^2\left(\frac{2 \pi f T}{2}\right)\right].\label{eqn:alpha}
\end{equation}

Also note that the normalised power per logarithmic frequency interval, $\Omega_{gw}(f)$, can also be written in terms of $S_h(f)$ and the characteristic strain spectrum (see equation \ref{d2dfvOmega}):
\begin{eqnarray}
\Omega_{gw}(f) &=& \frac{2 \pi^2}{3 H_0^2} f^3 S_h(f)\\
               &=& \frac{2 \pi^2}{3 H_0^2} f^2 h_c^2(f)\\
               &=& \frac{2 \pi^2}{3 H_0^2} A_g^2 f_{\rm 1yr}^2\left(\frac{f}{f_{\rm 1yr}}\right)^{2 \alpha + 2}.
\end{eqnarray}

\section{Simulated PPTA data sets}\label{sec:electronicTables}

An electronic supplement to this paper includes the simulated PPTA data sets.  Three tables are provided giving two-weekly sampling of the 20 PPTA pulsars with the addition of 100\,ns white, Gaussian noise.   The first table has no additional GW signal, the second table includes a GW background where $A_g = 10^{-14}$ and $\alpha = -2/3$ and the third table has a background where $A_g = 10^{-15}$.  The first column in the online tables gives the MJD of the simulated observation and the remaining 20 columns give the timing residuals for each of the 20 PPTA pulsars in the following order: PSRs~J0437$-$4715, J0613$-$0200, J0711$-$6830, J1022$+$1001, J1024$-$0719, J1045$-$4509, J1600$-$3053, J1603$-$7202, J1643$-$1224, J1713$+$0747, J1730$-$2304, J1732$-$5049, J1744$-$1134, J1824$-$2452, J1857$+$0943, J1909$-$3744, J1939$+$2134, J2124$-$3358, J2129$-$5721 and J2145$-$0750.

\end{document}